\documentclass[twocolumn,showkeys,prd,nofootinbib,floatfix]{revtex4-1}

\usepackage{amsfonts,amsmath,amssymb} 
\usepackage{graphicx,graphics,color}
\usepackage{gensymb}

\def\mnras{Mon. Not. Roy. Astron. Soc.} 

\def\mnras{MNRAS} \def\apj{ApJ} \def\apjl{ApJ} 
  \def\aap{A\&A} 
  
  \def\prd{Phys. Rev. D}
  
 \def\nat{Nature} \def\physrep{Phys. Rep.}
\def\pasj{PASJ}  

        \def\aj{Astron. J.}
\def\araa{annurev}

\begin{document}

\title{Dark-matter dynamical friction versus gravitational-wave emission in the evolution of compact-star binaries}

\author{L.~Gabriel G\'omez$^{1,2,3}$ and J.~A.~Rueda,$^{1,3,4}$}

\affiliation{$^1$Dipartimento di Fisica and ICRA, Sapienza Universit\`a di Roma, P.le Aldo Moro 5, I--00185 Rome, Italy}
\affiliation{$^2$University of Nice-Sophia Antipolis, 28 Av. de Valrose, 06103 Nice Cedex 2, France}
\affiliation{$^3$ICRANet, Piazza della Repubblica 10, I--65122 Pescara, Italy}
\affiliation{$^4$ICRANet-Rio, CBPF, Rua Dr. Xavier Sigaud 150, Rio de Janeiro, RJ, 22290--180, Brazil}

\date{\today}

\begin{abstract}
The measured orbital period decay of relativistic compact-star binaries, with characteristic orbital periods $\sim 0.1$~days, is explained with very high precision by the gravitational wave (GW) emission of an inspiraling binary in vacuum predicted by general relativity. However, the binary gravitational binding energy is also affected by an usually neglected phenomenon, namely the dark matter dynamical friction (DMDF) produced by the interaction of the binary components with their respective DM gravitational wakes. Therefore, the inclusion of the DMDF might lead to a binary evolution which is different from a purely GW-driven one. The entity of this effect depends on the orbital period and on the local value of the DM density, hence on the position of the binary in the Galaxy. We evaluate the DMDF produced by three different DM profiles: the Navarro-Frenk-White (NFW) profile, the non-singular-isothermal-sphere (NSIS) and the Ruffini-Arg\"uelles-Rueda (RAR) DM profile based on self-gravitating keV fermions. We first show that indeed, due to their Galactic position, the GW emission dominates over the DMDF in the NS-NS, NS-WD and WD-WD binaries for which measurements of the orbital decay exist. Then, we evaluate the conditions (i.e. orbital period and Galactic location) under which the effect of DMDF on the binary evolution becomes comparable to, or overcomes, the one of the GW emission. We find that, for instance for $1.3$--$0.2$ $M_\odot$ NS-WD, $1.3$--$1.3$~$M_\odot$ NS-NS, and $0.25$--$0.50$~$M_\odot$ WD-WD, located at 0.1~kpc, this occurs at orbital periods around 20--30 days in a NFW profile while, in a RAR profile, it occurs at about 100 days. For closer distances to the Galactic center, the DMDF effect increases and the above critical orbital periods become interestingly shorter. Finally, we also analyze the system parameters (for all the DM profiles) for which DMDF leads to an orbital widening instead of orbital decay. All the above imply that a direct/indirect observational verification of this effect in compact-star binaries might put strong constraints on the nature of DM and its Galactic distribution. 
\end{abstract}

\pacs{Valid PACS appear here}
\maketitle


\section{Introduction}

Compact-star binaries composed of neutron stars (NSs) and/or white dwarfs (WDs) have turned out to be rich laboratories of physics and astrophysics that allow to test fundamental theoretical predictions. In particular, NS-NS binaries have served to prove the existence of gravitational waves (GWs) \cite{1975ApJ...195L..51H} and the motion of matter and photons in the strong gravitational fields \cite{2006Sci...314...97K}, as well as other phenomena \cite{2012hpa..book.....L}. These latter aspects are of special interest in tests of general relativity and alternative theories of gravity \cite{2006Sci...314...97K,2013Sci...340..448A}. 

The orbital motion of such systems also offers the possibility of analyzing further effects. An interesting physical situation arises when the orbiting object moves through an extended medium which is formed, for instance, from the mass loss of the binary companion. This interaction can be thought as a drag force exerted by the circumbinary medium on the object in question, perturbing thereby its Keplerian orbital motion \cite{1976ApJ...204..879A}. This dynamical friction produced by the gravitational drag-force has been also studied in the context of different astrophysical phenomena such as mergers of star clusters, galaxies, and even galaxy clusters, to the inspiral of dwarf galaxies within dark-matter halos and the orbital evolution of massive black hole (BH) binaries in a stellar medium \cite{2008gady.book.....B}. Thus, dynamical friction plays an important role in the orbital evolution of many astrophysical systems. In a pioneering work, S.~Chandrasekhar \cite{1943ApJ....97..255C} calculated the dynamical friction force on a massive object traversing an infinite homogeneous collisionless background (representing the surrounding star neighbors).


It is thus natural to expect that a binary system moving through the galaxy can also experience a dynamical friction caused by collisionless DM particles, namely DM dynamical friction (hereafter DMDF), particularly in DM-dominated regions, as at the outer part of the Galactic halo and near the Galactic center \cite{2013PhR...531....1S}. The perturbed orbital motion may lead thus to interesting observable effects in the secular evolution of the orbital period. An interesting proposal was advanced in Ref.~\cite{2015PhRvD..92l3530P} on the possibility of inferring constraints to the DM density by determining the above DM effect on the orbital motion of binaries (see also the pioneering work by Bekenstein \& Zamir \cite{1990ApJ...359..427B}, for a  general discussion of collisionless background types as well as in the context of DM). They showed that the change in the orbital period could be due to the dynamical friction force exerted by the DM background on the binary. In that work, this effect was used to put an upper bound on the DM density in a given location of the Galaxy, independently of the density profile or the nature of the DM particles. It can be shown, however, that this upper limit is indeed fulfilled by any DM density profile consistent with the outer halo properties of the Milky Way. Thus, we explore in this work the dependence of the orbital period decay by DMDF on the different binary parameters and also on the DM density profile, in order to identify all possible physical situations suitable for an observational verification of the DMDF effect. For doing this we obtain DM profiles fulfilling definite Galactic-halo observables such as the escape velocity, the velocity dispersion and the one-halo scale length parameters. The velocity distribution function and the DM density profile are, as we shall show below, crucial elements in the dynamical friction force estimation. 

It is known that the DM in the outer part of our Galaxy is well described by a classical Maxwell-Boltzmann distribution, e.g. by a non-singular isothermal (hereafter NSIS) profile \cite{2008gady.book.....B}. However, depending on the DM nature (e.g. particle type), the DM density distribution can deviate from the classical Maxwell-Boltzmann behavior towards the inner regions of the Galaxy. This implies that the DMDF effect will depend according to the phase-space density consistent with the DM particle nature. We shall consider, for the sake of comparison, three DM models: 1) the NSIS profile, 2) the Navarro-Frenk-White (NFW) profile \cite{1996ApJ...462..563N}, and 3) the recently introduced Ruffini-Arg\"uelles-Rueda (RAR) model \cite{2015MNRAS.451..622R,2016arXiv160607040A}.


The RAR model is based on a self-gravitating system of massive (keV) fermions in thermodynamic equilibrium. The density profile of the RAR model exhibits a core-halo structure which allows to explain the DM distribution in galactic halos from dwarfs to big spirals, and predicts at the same time the presence of a DM high density core \cite{2015MNRAS.451..622R}. Under this approach and following the more realistic distribution function including violent relaxation processes \citep{2004PhyA..332...89C} and the escape velocity of particles, the Fermi-Dirac distribution function was subsequently introduced to describe the finite size of halos. In the case of the Milky Way, such a DM core can explain the observed dynamics near the Galactic center Sgr A* without invoking a central supermassive black hole for fermion masses in the range 48~keV $\lesssim m c^2 \lesssim 345$~keV \cite{2016arXiv160607040A} \footnote{See also Ref.~\cite{2016PhRvD..94l3004G} for the gravitational lensing properties of the RAR profile.}. 


Having established the DM density profiles we shall analyze, we now describe the structure of this work. We start by discussing in section \ref{sec:2} the effects which are commonly assumed to produce a change of the orbital period of binaries, putting special attention evidently to the one produced by GW emission. We analyze in section \ref{sec:3} the dynamical friction force and its main ingredients for the case when it is produced by DM and when it acts on binary systems. We analyze in section \ref{sec:4} the perturbation effect of DMDF on the orbital motion of the pulsar and reproduce some general results presented in \cite{2015PhRvD..92l3530P}. Furthermore, we introduce Galactic-halo observables in order to generalize the prescription presented in \cite{2015PhRvD..92l3530P} and present thus a more realistic estimation of dynamical friction effects. Finally, we present in section \ref{sec:5} the numerical results of $\dot{P}_{b}$ as a function of the radial position, the DM wind velocity and the orbital period. This latter computation leads us to compare directly the $\dot{P}_{b}$ due to GW emission to that given by DMDF. In section \ref{sec:5} we summarize our results and present a general discussion.

\section{Binary systems and orbital period decay by gravitational waves}\label{sec:2}

The precise pulsar timing measurements allow us to detect, with a high accuracy, tiny orbital effects which thus require a precise theoretical description of the orbital motion \cite{1975ApJ...195L..51H}. In the weak field regime (Newtonian approach), the binary motion of pulsar is simply described by the Kepler laws. However, relativistic and strong-field effects in the orbital motion should be taken into account in the vicinity of a close-orbit binary pulsar \cite{2006Sci...314...97K}. These relativistic effects can be described, for the known binaries, with sufficient accuracy in terms of the called \emph{post-Keplerian} parameters that account for departures from Newtonian Keplerian dynamics owing e.g. to the GW emission, time delay caused by the curvature of space-time near the pulsar (Shapiro delay), and relativistic time dilation \cite{2014grav.book.....P}. There exists a variety of effects that affect the orbital period stability and they can be, roughly speaking, classified in two large groups: kinematic and intrinsic to the system. The former include the effects of a secular increase due to the Galactic gravitational potential, secular acceleration resulting from the pulsar’s transverse velocity (proper motion of the pulsar) and the cluster’s gravitational field; while the latter is related to ``local'' effects in the system as mass loss either from the pulsar or its companion and the GW emission among others\footnote{For a more detail description of possible effects on the observed period decay see Refs.~\cite{1991ApJ...366..501D,2012hpa..book.....L}.}. After subtracting kinematic effects from the observed change of the orbital period, the remaining intrinsic period decay has been shown to be explained by the GW emission predicted by general relativity of an inspiraling binary in vacuum.
 
The orbital period decay owing to the GW emission of a binary spiraling in circular orbits is given by
\begin{equation}
\dot{P}^{\rm GW}_b=-\frac{192\pi}{5} \left(\frac{2\pi G \mathcal{M}}{c^3 P_b}\right)^{5/3},\label{c9}
\end{equation}
where $G$ is the gravitational constant, $P_b$ is the orbital period, $\mathcal{M} = (m_p m_c)^{3/5}M^{-1/5}$ is the so-called chirp mass and $M=m_p+m_c$ is the total mass, with the subscripts $p$ and $c$ denoting the primary component and its companion, respectively.

The theoretical prediction of general relativity given by Eq.~(\ref{c9}) was first verified with the observed intrinsic orbital period decay of the famous Hulse-Taylor binary pulsar PSR B1913+16 \cite{1975ApJ...195L..51H}, which is explained with an accuracy of 99,8\%. Later on, additional successful verifications in other relativistic NS-NS and NS-WD binaries have been made and with even higher accuracy. We refer the reader to Ref.~\cite{2015ASSP...40....1A} for a review on this subject and also Table~\ref{tab:pulsars}.

As we have mentioned, the above orbital period decay by GW emission is calculated under the assumption of binary motion in empty space. We shall explore below the effect of the presence of DM background on the orbital motion via dynamical friction, i.e. by DM gravitational drag. We shall infer the predicted orbital period time derivative by this phenomenon to then compare it with the one produced by the GW emission.

\section{Dynamical friction force and its main ingredients}\label{sec:3}

Dynamical friction has been widely used to account for the drag force when an object is moving through a collisionless medium of field particles. This drag induces a wake of medium particles on the object with a characteristic overdensity proportional to its mass \cite{2008gady.book.....B}. In his seminal work, Chandrasekhar \cite{1943ApJ....97..255C} computed the dynamical friction force onto an object that move in an infinite homogeneous stellar medium obeying a Maxwell-Boltzmann velocity distribution, taking into account only the contribution of the field particle velocities smaller with respect to the object's velocity. However, the dynamical evolution of many astrophysical systems is driven by dynamical friction in a more realistic way \cite{2008gady.book.....B}. We consider the drag force, $\textbf{f}_{fr,i}$, experienced by a test body of mass $m_i\gg m$, being $m$ the DM particle mass, and with orbital velocity $v_i$ moving through the DM background with velocity distribution function $f(u)$ \cite{1943ApJ....97..255C,2008gady.book.....B}:
\begin{align} 
\textbf{f}_{fr,i}&=-4\pi G^2 m_i^2 m \left( \int_0^{\tilde{v}_i} d^3u f(u) \ln\left[\frac{b_{\rm max}}{G m_i} (\tilde{v}_i^{2}-u^2)\right] \right. \nonumber\\
& + \left. \int_{\tilde{v}_i}^{v_{\rm esc}} d^3u f(u) \left[\ln\left(\frac{u+\tilde{v}_i}{u-\tilde{v}_i}\right)-2\frac{\tilde{v}_i}{u}\right]\right)\frac{\tilde{\textbf{v}}_i}{\tilde{v}_i^{3}},\label{a01}
\end{align} 
where the integral in the first term, accounts for low velocity contributions (fraction of particles moving slower than the object) while the integral in the second term, refer to the faster particles, limited by the escape velocity $v_{\rm esc}$ according to the Galactic gravitational potential.\footnote{It has been recently shown that the incorporation of the tidal radius into the background system can produce interesting features in infalling satellites in large cored galaxies \cite{2016MNRAS.463..858P}.} $b_{\rm max}$ is the maximum impact parameter defined below in Eq.~(\ref{eq:bmax}). The above equation takes into account the orbital velocity of each object with respect to the DM wind relative to the center of mass of the binary system: $\tilde{\textbf{v}}_{i}=\textbf{v}_{i}+\textbf{v}_{w}$, with $\textbf{v}_{w}=v_{w}(\cos{\alpha} \sin{\beta},\sin{\alpha}\sin{\beta},\cos{\beta})$ and $\beta$ and $\alpha$ being the angles between the wind velocity vector and the perpendicular axis of the binary orbital plane and the projection of the wind velocity vector with an axes lying in the orbital plane, respectively. There are at least two different cases of wind velocities: bound and unbound binaries to the galaxy potential. In the former the DM wind velocity can be assumed as the negative of the binary circular velocity with respect to the galactic center $\textbf{v}_{w} = -\textbf{v}_{\rm rot}$. The latter case occurs often in binaries with NS components in which the system received a high kick velocity from the supernova event \cite{1994Natur.369..127L}. For high kick velocities the binary circular velocity with respect to the galactic center can be neglected \cite{2006LRR.....9....6P} and we can assume $\textbf{v}_{w} = -\textbf{v}_{T}$, where $\textbf{v}_{T}$ is the transversal velocity of the system. For intermediate kicks, the system can remain bound and we can consider, in a more general  case, $\textbf{v}_{w}=\textbf{v}_{\rm rot}+\textbf{v}_{T}$. Thus, we shall consider the value of $v_w$ as a free parameter that can assume values ranging from 10~km~s$^{-1}$ all the way to 1000~km~s$^{-1}$ following the above discussion. There is the additional possibility for the binary components to experience an intrinsic DM wind. However, up to the best of our knowledge, there is no observational evidence of an intrinsic rotation of the DM with respect to the Galactic center and thus we do not consider it in our estimates.

It is important also to mention that the condition $L/a\ll 1$, where $L$ is the size of the component's wake and $a$ the orbital separation, must be fulfilled in order that Eq.~(\ref{a01}) becomes linearly applicable to each binary component \cite{1990ApJ...359..427B,2008gady.book.....B}. Since $L$ is of the order of the radius of the sphere of gravitational influence of each component -- see Eq.~(\ref{eq:RA}) below --  this means that we are limited to binary systems with orbital velocities smaller than the velocity dispersion of the DM background. Namely, we deal with binary systems with sufficiently large orbital periods (small orbital compactness) so that each binary component does not interact with its respective companion's wake. Furthermore, we treat the binary system as composed of point masses no matter their internal structure. Thus, we can apply this approach under the above conditions to binary systems such as NS-NS/NS-WD \cite{2015ASSP...40....1A} and WD-WD \cite{2012ApJ...757L..21H}, or any other possible binary system of astrophysical interest.

%
%
We proceed now to introduce the most relevant ingredients entering into the computation of the dynamical friction force on the binary system. This analysis allow us to establish more accurately our system in terms of Milky Way galactic observables and to characterize more realistically the DM density properties.
%
\subsection{The Coulomb logarithm} \label{sec:3A}

%
\begin{figure} 
\centering 
\includegraphics[width=\hsize,clip]{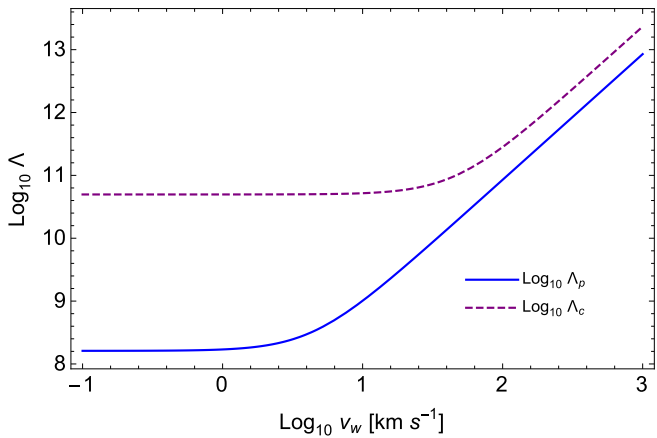}
\caption{Coulomb logarithm for the primary, $\log_{10}\Lambda_{p}$ (blue line), and for the secondary, $\log_{10}\Lambda_{c}$ (red line), as a function of the DM wind velocity. The primary is a NS of $1.3~M_\odot$ and the companion secondary is a WD of $0.2~M_\odot$. The NS-WD binary has an orbital period $P_{b}=100$~days and $\beta=\pi/2$. The differences between the Coulomb logarithms lead every component of the system to experience distinct gravitational interactions with its respective wake.} \label{lambda-vw}
\end{figure}

The Coulomb logarithm in the Chandrasekhar's dynamical friction formula accounts for the finite size of the system and is defined as the ratio of the maximum and minimum impact parameters for encounters, respectively $b_{\rm max}$ and $b_{\rm min}$, i.e.
\begin{equation}
\log{\Lambda}\equiv\log{\left(\frac{b_{\rm max}}{b_{\rm min}}\right)}.\label{a03}
\end{equation}
It is assumed typically that $b_{\rm max}$ is of the order of the size of the system, and $b_{\rm min}$ is defined as the impact parameter for a 90$^\circ$ deflection \cite{2008gady.book.....B}
\begin{equation}\label{eq:bmax}
b_{\rm max}\approx a, \qquad b_{\rm min}={\rm max}(r_h,R_A),
\end{equation}
where $b_{\rm max}$ can be taken as the effective size of the system (the binary orbital separation) and $r_h$ is the half-mass radius of the subject system. This is the radius that contains the body's half-mass and should be taken as $b_{\rm min}$ in the case it be an extended body. However it does not correspond to the present case. We instead adopt $b_{\rm min}=R_{A}$, where $R_A$ is the radius at which a particle of the surrounding medium is affected by the sphere of gravitational influence of the test body, namely:
\begin{equation}\label{eq:RA}
R_{A,i}=\frac{G m_i}{\tilde{v}_{i}^2},
\end{equation}
being $\tilde{v}_i$ the relative velocity of the object with respect to the DM wind velocity as we defined above. 

We can see from here that dynamical friction force is determined by the local distribution of matter producing the wake around each object. This also establishes the characteristic size of the wake. It is here assumed that $b_{\rm max}\gg b_{\rm min}$ and $b_{\rm max}$ is set to be the length scale over which the density can be assumed to be constant for a given system at fixed radial position. It is important to note that the choices of the impact parameters are somewhat arbitrary. However, we guarantee that the condition $\Lambda\gg 1$ is satisfied. 

As an example we plot in Fig.~\ref{lambda-vw} the Coulomb logarithm as a function of the wind velocity for a $1.3+0.2~M_\odot$ NS-WD binary with orbital period $P_{b}=100$~days. We stress that the Coulomb logarithm does not change with $\beta$, we choose however $\beta\neq 0$ to perform, in a more general way, a study of the orbital period decay. We also note that, for $v_w\ge 80$~km~s$^{-1}$ there are not large differences between the Coulomb logarithm for each object. However, we will take into account these small differences for accuracy even though we consider, in some cases, large values for the wind velocity.

\subsection{Velocity distribution function}\label{sec:3B}

The evolution of a collisionless self-gravitating system is determined by the Vlasov-Poisson equation that sets the conservation of the phase-space density \cite{2008gady.book.....B}. This distribution function fully specifies the dynamic of a colissionless system. For instance, for spherical systems, the mass density is proportional to $\int d^{3}v f$. It is also possible to derive the distribution function of a collisionless system for a given self-consistent density profile $\rho$ following the Eddington's formula \cite{1916MNRAS..76..572E}
\begin{equation}
f(\mathcal{E})= \frac{1}{\sqrt{8}\pi^{2}}\left[\int_{0}^{\mathcal{E}} \frac{d^{2}\rho}{d\Psi^{2}}\frac{d\Psi}{\sqrt{\mathcal{E}-\Psi}}+\frac{1}{\sqrt{\mathcal{E}}} \left( \frac{d\rho}{d\Psi}\right)_{\Psi=0}\right],\label{a4}
\end{equation}  
where we have introduced the relative potential and binding energy (per unit mass) defined respectively as: $\Psi=-\Phi+\Phi_{0}$ and $\mathcal{E}=-E+\Phi_{0}=\Psi-\frac{1}{2}v^{2}$. For a spherical system with an isotropic velocity dispersion, the phase space distribution function of dark halos depends only on the energy and not on the angular momentum. The above formula is particularly useful when we seek for a distribution function to associate with a density profile obtained from other methods. We shall apply this procedure in the appendix, to the NSIS and to the phenomenological NFW profile to validate the approximation of considering, within our estimations, the Maxwell-Boltzmann distribution for these both profiles. Significant but very small differences appear between the Maxwell-Boltzmann distribution and the distribution functions associated to the NFW and NSIS profile at nearby unbound energies, as can be seen in Fig.~\ref{Eddign_sol}. In addition, the unbound energy ($\mathcal{E}=0$) permits, in fact, to define the escape velocity $v_{esc}=\sqrt{2|\Psi|}$. We shall see that the contribution of particles moving faster than the object and limited by the escape velocity, do not contribute substantially to the dynamical friction force. This consequence supports the fact of considering the Maxwell-Boltzmann distribution to describe the velocity distribution for the aforementioned profiles. The main motivation of this approach is then, due to the numerical facilities that the simple Maxwell-Boltzmann distribution provides in the computation of the dynamical friction force.  


Accordingly, for the sake of comparison, let us assume then that the virialized NFW and NSIS halos, follow the Maxwell-Boltzmann distribution function
\begin{equation}
f^{\rm MB}(u)= \frac{n_{0}}{(2\pi\sigma^{2})^{3/2}}\exp\left(-\frac{u^{2}}{2\sigma^{2}}\right),\label{f0}
\end{equation}
where $n_{0}$ is the particle number such that $\rho=n_{0}m$ and $\sigma$ is the velocity dispersion which is defined in terms of the DM gravitational potential through the Jeans Eq.~(\ref{a5}).

For the RAR model, we consider self-consistently a Fermi-Dirac distribution function with energy cutoff $\epsilon_{c}$ to describe the velocity distribution of self-gravitating halos in thermodynamic equilibrium  \cite{2015MNRAS.451..622R}\footnote{See also Ref.~\cite{2004PhyA..332...89C} for a general discussion about the conditions under which statistical equilibrium state is reached.}
\begin{equation} 
f_{c}(p)= \frac{g m^{3}}{h^{3}}\left\{ \begin{array}{ll}\frac{1-e^{(\epsilon-\epsilon_{c})/kT}}{e^{(\epsilon-\mu)/kT}+1}  &
\epsilon\leqslant \epsilon_{c}, \\ 0 &  \epsilon >\epsilon_{c}. \end{array}
\right. \label{f1}\end{equation}
Here $\epsilon=\sqrt{c^{2}p^{2}+m^{2}c^{4}}-mc^{2}$ is the particle kinetic
energy, $m$ is the particle mass, $\mu$ is the chemical potential (with the
particle rest mass subtracted off), $T$ is the temperature, $k$ is the
Boltzmann constant. The quantity $g$ denotes as usual the particle spin degeneracy ($g=2$ in our case) and $h$ is the Planck constant. It is important to stress that, the parameter $\epsilon_{c}$ serves to account for the finite size of galaxies. Note also that for $\epsilon_{c}\rightarrow +\infty$, we recover the Fermi-Dirac distribution. In the non-degenerate limit $\mu\rightarrow -\infty$, we recover on the other hand the classical King model  \cite{1965AJ.....70..376K}, which reduces to the Boltzmann distribution in the limit  $\epsilon_{c}\rightarrow +\infty$.

As we have mentioned, we are going to explore in this work the dynamical friction force effects on binary systems produced by DM profiles. However, it is important to note that, the dynamical friction force depends actually on the velocity distribution function whereby the introduction of the DM density profile, is somehow artificial; but in any case, it should be self-consistent for a given velocity distribution function according to the previous discussion.


\subsection{The escape velocity}\label{sec:3C}

The escape velocity is defined in terms of the gravitational potential $\phi(r)$ of the background\footnote{Note that we are ignoring the gravitational potential produced by the binary system as well as other possibles contributions, as those produced by the baryonic component.} as $v_{\rm esc}=\sqrt{-2\phi}$. The latter can be determined completely at any radius scale for a given density profile as follows
\begin{equation}
\phi(r)=4\pi G \left[\frac{1}{r} \int_{0}^{r}dr' r'^{2} \rho(r')+\int_{r}^{\infty} dr' r' \rho(r') \right].\label{a02}
\end{equation}
The observed escape velocity of the Milky Way (considering the Galactic components, disk, bulge and halo) was found to be in the range $498$~km~s$^{-1}$ $\lesssim v_{\rm esc}\lesssim 608$~km~s$^{-1}$ at the solar position, at $90\%$ confidence interval and median likelihood of $544$ km~s$^{-1}$ \cite{2007MNRAS.379..755S}. The RAVE survey has recently found the local escape speed to be $v_{\rm esc}=533^{+54}_{-41}$~km~s$^{-1}$ \cite{2014A&A...562A..91P}. These values depend significantly on the mass exterior to the solar circle within a certain halo radius $r_h$. For example, the halo mass $M_{DM}(r_h = 40$~kpc$) \sim 2\times 10^{11}M_{\odot}$ is consistent with the dynamics of the outer DM halo as was recently indicated in \cite{2014MNRAS.445.3788G}. We note therefore that the Galactic escape velocity is either lower or closely equal to the orbital velocity of the binary pulsar for periods around $P_b\approx 0.1$~days. For large orbital periods $P_b\approx 100$~days, the orbital velocity is always well below the escape velocity. These two facts imply therefore that the contribution of the second integral (fast particles) to the dynamical friction force could be very small in most cases but not negligible in general. We will keep this term for a general study since, as we will see, it also leads to a change of sign in the orbital period time derivative (i.e. from decay to widening) for some values of the period as well as for the DM wind velocity. 

\subsection{The density profile}\label{sec:3D}

\subsubsection{The NFW profile}

We first recall the widely used phenomenological DM density profile arising
within the $\Lambda$CDM cosmological paradigm, i.e. the NFW profile
\cite{1996ApJ...462..563N}
\begin{equation}
\rho(r)=\frac{\rho_{c}}{(r/r_{s})(1+r/r_{s})^{2}},\label{a1} \end{equation}
where $\rho_c$ is the characteristic density and $r_s$ is the scale
radius. This density profile exhibits a sharp cusp in the inner region
$\rho\varpropto r^{-1}$ while in the halo part the density scales as
$\rho\varpropto r^{-3}$.

It is worth to mention that there is an active debate in the literature on which is the best representation of the DM density profile that originates from the $\Lambda$CDM paradigm. For instance, some simulations has pointed out that the density profile of DM halos might be actually shallower than the one given by the NFW profile and found a cored structure represented more accurately by an Einasto profile (see Ref.~\cite{2006AJ....132.2685M} for details).  It is out of the scope of this work to make an assessment on this issue and thus, for the sake of example, we adopt the NFW profile as the DM profile associated with the $\Lambda$CDM scenario. As we shall see, since the NSIS and the RAR profiles show also a shallower, cored inner halo\footnote{The similarity between the Einasto profile and the RAR profile in the inner halo region has been shown in Ref.~\cite{2015ARep...59..656S}.}, they are useful to analyze the differences that arise in the DMDF effect between cuspy and cored density profiles.

\subsubsection{The NSIS profile}

Another often adopted DM density profile which also yields the asymptotic flatness of the rotation curves is represented by the NSIS profile \cite{1999MNRAS.307..203S}:
\begin{equation} \rho(r)=\frac{\rho_{0}}{1+(r/r_{0})^{2}},\label{a2}
\end{equation}
where $\rho_{0}$ is the central density and $r_{0}$ is the core radius.

\subsubsection{The RAR profile}

\begin{figure} 
\centering 
\includegraphics[width=\hsize,clip]{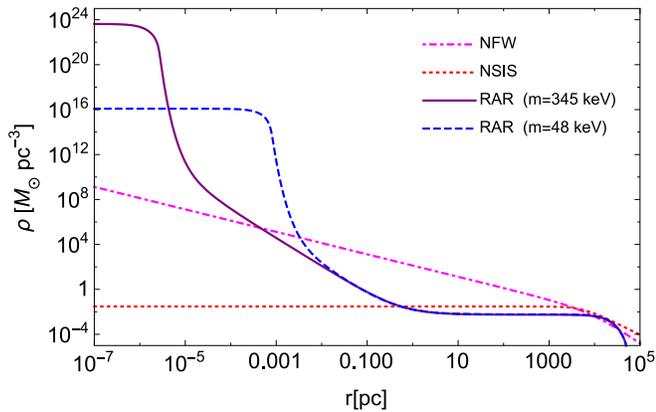}
\caption{Distribution of DM in MW-type galaxies predicted by the RAR model. The solid line in the legend, refers to the most compact solution for $m=345$~keV. For comparison we show, with the dashed blue line, the  solution for $m=48$~keV. There are also shown the NFW and NSIS profiles given by Eqs.~(\ref{a1}) and (\ref{a2}), respectively. The free parameters in these profiles were taken from \cite{2013PASJ...65..118S} and \cite{2009PASJ...61..227S}, respectively, satisfying the
same (total) rotation curve data as in the RAR case, with the corresponding
considerations of bulge and disk counterparts.} \label{density}
\end{figure}

We will also examine the DMDF in the case of the RAR model \cite{2015MNRAS.451..622R,2016arXiv160607040A}. This model describes the DM distribution along the entire galaxy in a continuous way, i.e. from the halo
part to the Galactic center and without spoiling the baryonic component which dominates
at intermediate scales. 
Likewise, the density $\rho$ and pressure $P$ for the Fermi-Dirac distribution function are defined respectively by
\begin{eqnarray} \rho&=&\frac{g}{h^{3}}m \int_{0}^{\epsilon_{c}} f_{c}(p)
\left(1+\frac{\epsilon(p)}{m c^{2}}\right)d^{3}p,\label{b4}\\
P&=&\frac{2}{3}\frac{g}{h^{3}} \int_{0}^{\epsilon_{c}} f_{c}(p)
\frac{1+\epsilon(p)/2mc^2}{1+\epsilon(p)/mc^{2}}d^{3}p,\label{b5}
\end{eqnarray}

Assuming a self-gravitating system of massive fermions (within the standard
Fermi-Dirac phase-space distribution) in thermodynamic equilibrium, the DM
density profile was computed in \cite{2015MNRAS.451..622R}. By imposing
fixed boundary conditions at the halo and including the fulfillment of the
rotation curves data, the parameters of the system have been constrained.
This procedure was applied for different types of galaxies from dwarfs to
big spirals exhibiting a universal \emph{compact core} - \emph{diluted
halo} density profile. An extended version of the RAR model was recently
presented \cite{2016arXiv160607040A}, by introducing a fermion energy
cutoff $\epsilon_{c}$ in the fermion distribution. Importantly, this generalization in the statistics naturally arises by studying the stationary solution of a generalized Kramer statistics which includes the effects of escape of particles and violent
relaxation \cite{2004PhyA..332...89C}. The new emerging density profile
serves to account for the finite galaxy sizes due to the more realistic
boundary conditions, while it opens the possibility to achieve a more
compact solution for the quantum core working as a good alternative to the BH scenario in Sgr A* (see, Ref.~\cite{2016arXiv160607040A}, for details).
The narrow particle mass range provides several solutions to
satisfy either the rotation curve data in the halo part or both sets of
data, namely including additionally the orbits of the S-cluster stars such
as the S2 star, necessary to establish the compactness of the DM central core. A comparison between the RAR model, NFW profile and NSIS for MW-like spiral galaxies is also shown in Fig.~\ref{density}, describing the outstanding inner structure below parsec scale for the RAR profile.

It is important to clarify that the above DM density profiles are obtained without considering a DM-baryonic matter feedback nor DM self-annihilations. As shown in Ref.~\cite{2005PhRvD..72j3502B}, these effects might produce changes in the DM density profile. We expect, however, the former to be important only locally in massive clusters and the latter stands on the largely model-dependent unknown DM nature. Thus, for the sake of generality, we shall not consider these effects in this work.

\subsection{The velocity dispersion}\label{sec:3e}
According to observations of stars in outer part of halos and numerical simulation, the stellar velocity dispersion of the Milky-Way halo $\sigma_{r}$, shows an almost constant value around $120$~km~s$^{-1}$ at scales of 20~kpc where DM is supposed to dominate and the circular velocity $V_c$ exhibits a flat behavior. Assuming that the galactic halo is stationary and spherically symmetric, it is possible to derive the DM radial velocity dispersion from the Jeans equation\footnote{Note that it is not the (observed) line of sight velocity dispersion of tracers.}
\begin{equation}
\frac{1}{\rho(r)} \frac{d(\rho(r)\sigma_{r}^{2})}{dr}+2\frac{\beta\sigma_{r}^{2}}{r}=-\frac{d\phi(r)}{dr}=-V_{c}^{2},\label{a5}
\end{equation}
where $\beta=1-\sigma_{\theta}^{2}/\sigma_{r}^{2}$ is the velocity anisotropic parameter, that in the isotropic case, takes evidently the value $\beta=0$.

The circular velocity $v_c$ is defined  by the local radial gradient of the potential while the radial velocity dispersion $\sigma(r)$ depends on the shape of the potential at exterior radii. For a non-rotating spherical system the relation between these quantities is given by
\begin{equation}
v_{c}^{2}=-\sigma_{r}^{2}\left( \frac{d\ln\rho}{d\ln r}+\frac{d\ln\sigma_{r}^{2}}{d\ln r}+2\beta\right),
\end{equation}
where the first term in parenthesis is (minus) the logarithmic slope $\gamma$ of the density profile. For the singular isothermal sphere with a Maxwell Boltzmann distribution, the simple relation $v_{c}^{2}=2\sigma_{r}^{2}$ is satisfied for all radii. We note that, instead, for the NFW profile one obtains $1\leq\gamma\leq 3$ and hence, this simple relation between the circular velocity and the velocity dispersion is not fulfilled at all radii (except at the virial radius where $\gamma=2$, see e.g. Ref.~\cite{2010MNRAS.402...21N}). Therefore, in order to find the right velocity dispersion profile for a given density profile, with associated gravitational potential, we solve hence the Jean equation for the isotropic case along the entire the Galaxy.


\section{Orbital period evolution}\label{sec:4}

In this section we study the DMDF effect as an intrinsic effect on the binary system motion. Hence, in order to analyze the perturbed Keplerian orbit of binary systems, we use the osculating formalism that permits to obtain the  sequence of perturbed orbits \cite{2014grav.book.....P}. We follow particularly both the formulation and the derived analysis presented in \cite{2015PhRvD..92l3530P} to compute the orbital period decay due to DMDF. We start by defining the relative acceleration between two bodies as 
\begin{equation}
\dot{\textbf{v}}=-\frac{GM}{r^{3}}\textbf{r}+\textbf{f},\label{c1}
\end{equation}
with $\textbf{f}=a_{1}\eta \textbf{v}+a_{2} \textbf{v}_{w}$ for the case in which the perturbing force is taking to be the drag force measured on the center of mass. To zeroth order, the orbital velocity obeys a Keplerian motion, $v=\Omega_{0}r_{0}$, with $\Omega_{0}$ and $r_{0}$ being the angular velocity and orbital separation, respectively. We have also introduced the definitions: $\eta=\mu/M$, $\mu= m_{p}m_{c}/M$ and $M=m_{p}+m_{c}$. From here, the perturbed orbital elements can be then written as follows
\begin{equation}
\dot{a}=2\sqrt{\frac {r_{0}^{3}}{GM}} S(t),\label{c2}
\end{equation}
\begin{equation}
\dot{e}=2\sqrt{\frac {r_{0}}{GM}} [R(t)\sin(\Omega_{0}t)+2S(t)\cos(\Omega_{0}t)],\label{c3}
\end{equation}
\begin{equation}
\dot{i}=2\sqrt{\frac {r_{0}}{GM}} W(t) \cos(\Omega_{0}t+\omega),\label{c4}
\end{equation}
\begin{equation}
\dot{\Omega}=\frac{1}{\sin i}\sqrt{\frac {r_{0}}{GM}} W(t)\sin(\Omega_{0}t+\omega),\label{c5}
\end{equation}
where the orbital parameters $a$, $e$, $\omega$, $i$ and $\Omega$ are the semiaxis major, the eccentricity, the longitude of pericenter, the inclination and longitude of the ascending node, respectively. In the right side of Eqs.~(\ref{c2})-(\ref{c5}), the source terms $S(t)$, $R(t)$ and $W(t)$ have been defined as a functions of the dynamical friction force as well as the wind velocity vector according to \cite{2015PhRvD..92l3530P}
\begin{equation}
S(t)=a_{1}\eta v-a_{2} v_{w} \sin\beta \sin(\Omega_{0}t-\alpha),
\end{equation}
\begin{equation}
R(t)=a_{2}v_{w}\sin\beta \cos(\Omega_{0}t-\alpha),
\end{equation}
\begin{equation}
W(t)=a_{2}v_{w}\cos\beta.
\end{equation}
The rate of change of the separation with time leads consequently to a change of the orbital period $P_{b}=2\pi/\Omega_0$ given by \cite{1980stph.book.....L}
\begin{equation}
\frac{\dot{P_{b}}}{P_{b}}=\frac{3}{2}\frac{\dot{a}}{r_{0}}.\label{c6}
\end{equation}
This relation along with Eq.~(\ref{c2}) provide the time derivative of the orbital period\footnote{We note there is a typo in Eq.~(18) of Ref.~\cite{2015PhRvD..92l3530P}, namely when compared with Eq.~(\ref{c7}) of our present work it shows an extra factor $v/2$ which leads the equation to be dimensionally incorrect.}
\begin{equation}
\dot{P_{b}}(t)=3 P_{b}[a_{1}\eta-a_{2}\Gamma \sin\beta\sin(\Omega_{0}t-\alpha)].\label{c7}
\end{equation}
The resulting secular change in the orbital period is obtained by averaging over one period $P_{b}$, namely (see e.g. Ref.~\cite{2015PhRvD..92l3530P}):
\begin{equation}
\langle\dot{P}_{b}\rangle=\frac{1}{P_{b}}\int_{0}^{P_{b}} \dot{P_{b}}(t) dt.
\end{equation}
In the above formulation we have introduced the same definitions as in \cite{2015PhRvD..92l3530P} for an easier comparison of the results: $\Gamma=v_{w}/v$, $\Delta_{\pm}=\Delta\pm 1$, $\Delta=\sqrt{1-4\eta}$. The coefficients $a_i$ can be written in terms of the integral velocity contribution function
\begin{equation}\label{eq:bi}
b_i=\frac{1}{\rho(r)\log\Lambda_{i}}\frac{I_{i}}{\tilde{v}_{i}^{3}},
\end{equation}
as
\begin{equation}
a_{1}=-(A_{1}b_{1}+A_{2}b_{2}),\qquad a_{2}=\frac{1}{2}(A_{1} b_{1}\Delta_{+}+A_{2}b_{2}\Delta_{-}),\label{c8}
\end{equation}
with $A_{i}=4\pi\rho(r)\log\Lambda_{i}G^{2}M$. The definition of $b_i$ in the more general form given by Eq.~(\ref{eq:bi}) allows the use of any velocity distribution function, or equivalently any density profile through the integral term $I_i$ (term in parenthesis in Eq.~(\ref{a01})). This feature is contrary to the analyzed case in \cite{2015PhRvD..92l3530P} where the Maxwell-Boltzmann distribution function was only considered there.

It is clear that the initial phase $\alpha$ can be set to any value without loss of generality, hence we set $\alpha=0$ for simplicity. In the next section \ref{sec:5}, we compute then the secular change of $\dot{P}_b$ for different density profiles with velocity dispersion profile determined by Eq.~(\ref{a5}) and associated velocity distribution function described by Eqs.~(\ref{f0}) and (\ref{f1}). The incorporation of the radial scale dependence of these quantities, leads to reduce the number of free parameters presented in early calculations \cite{2015PhRvD..92l3530P}, as already pointed out previously. 

There may be other contributions to a secular change of the orbital period in addition to the DMDF and the gravitational-wave emission. A common effect in binaries with ordinary star components is the mass loss by star winds or accretion. A change of mass in the system would produce a change in the orbital period of the type $\dot{P}_b/P_b = -\dot{M}/M$, thus mass loss increases the orbital period (orbital widening) and mass accretion decreases it (orbital decay). In our case of binaries composed of compact stars the mass loss by winds is unlikely and accretion of matter from one component into the other could occur only via Roche lobe overflow for extremely short binary periods near the merging process. It remains the possibility of accretion of DM particles onto the binary components leading to a shrink of the orbit. The assessment of the importance of this effect, however, relies on the unknown cross-section between DM and baryonic matter inside the stars (see, e.g., Ref.~\cite{2013ApJ...774...48M}). Thus, for the sake of generality of our conclusions, we shall not include this effect in our estimates.

%
\section{Numerical results}\label{sec:5}

\begin{table*}
\begin{tabular}{llllllllll}
Name & Type & $m_{p}$~[$M_{\odot}$] & $m_{c}$~[$M_{\odot}$] & $P_{b}$~[days] &  d~[kpc] & $\dot{P}_{b}^{\rm int}$~[$10^{-12}$]  & $\dot{P}_{b}^{\rm GW}$~[$10^{-12}$] & $\dot{P}_{b,NFW}^{\rm DF}$~[$10^{-21}$] & $\dot{P}_{b,{\rm RAR}}^{\rm DF}$~[$10^{-21}$] \\ \hline
J0737-3039 & NS-NS & 1.3381(7) & 1.2489(7) & 0.104 & 1.15(22) & -1.252(17) & -1.24787(13) & -10.498 & -7.860\\ 
B1534+12 & NS-NS & 1.3330(4) & 1.3455(4) & 0.421 & 0.7 & -0.19244(5) & -0.1366(3) & -244.166 & -27.827\\ 
J1756-2251 & NS-NS & 1.312(17) & 1.258(17) & 0.321 & 2.5 & -0.21(3) & -0.22(1) & -0.271 & -20.695\\ 
J1906+0746 & NS-NS & 1.323(11) & 1.290(11) & 0.166 & 5.4 & -0.565(6) & -0.52(2) & -2.655 & -11.176\\ 
B1913+16 & NS-NS & 1.4398(2) & 1.3886(2) & 0.325 & 9.9 & -2.396(5) & -2.402531(14) & -7.942 & -17.747\\ 
B2127+11C\footnote{This binary is located in the globular cluster M15 \cite{1990Natur.346...42A}. However we have made here a simple estimation of the DMDF effect assuming that the DM local density in its location does not change abruptly within the globular cluster, which may not be the case. This point is better discussed in footnote \ref{footnote 1}. For a comprehensive list of all known binaries in globular clusters see http://www.naic.edu/~pfreire/GCpsr.html  and references therein.} & NS-NS & 1.358(10) & 1.354(10) & 0.333 & 10.3(4) & -3.961(2) & -3.95(13) & -8.083 & -17.0154 \\ 
J0348+0432 & NS-WD & 2.01(4) & 0.172(3) & 0.104 & 2.1(2) & -0.273(45) & -0.258(11) & -0.399 & -1.514\\ 
J0751+1807 & NS-WD & 1.26(14) & 0.13(2) & 0.263 & 2.0 & -0.031(14) & --- & -1.022 & -2.587\\ 
J1012+5307 & NS-WD & 1.64(22) & 0.16(2) & 0.60 & 0.836(80) & -0.15(15) & -0.11(2) & -3.404 & -7.343 \\ 
J1141-6545 & NS-WD & 1.27(1) & 1.02(1) & 0.20 & 3.7 & -0.401(25) & -0.403(25) & -3.578 & -11.469 \\ 
J1738+0333 & NS-WD & 1.46(6) & 0.181(7) & 0.354 & 1.47(10) & -0.0259(32) & -0.028(2) & -2.120 & -4.379\\ 
WDJ0651+2844 & WD-WD & 0.26(4) & 0.50(4) & 0.008 & 1 & -9.8(28) & -8.2(17) & -0.014 & -0.207 \\
\hline
\end{tabular}
\caption{Intrinsic orbital decays for several binary systems in the Galaxy as well as the ones predicted by GR and DM dynamical friction. There, it is also shown the values of mass binaries, orbital periods and distances measured from the Galactic center. This information is taken completely from Table \ref{tab:pulsars}. in Ref.~\cite{2015ASSP...40....1A} and references therein. For updated values of masses of neutron stars see \cite{2016ARA&A..54..401O}. We have simply added the last row for the WD-WD binary and the last two columns to show the orbital decay predicted by DMDF for the NFW profile and the RAR model.}\label{tab:pulsars}
\end{table*}

We present now the dependence of $\dot{P}_b$ according to Eq.~(\ref{c7}) on the free parameters: the orbital period, the DM wind velocity and the radial position of the binary measured from the Galactic center. 

Once the density profile has been chosen, and the binary position has been fixed, the velocity distribution function, the velocity dispersion, as well as the escape velocity that constrains the maximum velocity in phase space, (upper limit in the second integral Eq.~(\ref{a01})) can be determined uniquely. Thus, for an observed binary at a known galactic position, the above quantities acquire values that can not be treated as uncorrelated and fully free parameters.

In the following analysis we adopt for the RAR model the solution for the Milky Way with a particle mass $m=345$~keV, which has the density profile with the most compact quantum core  (see Fig.~\ref{density}). We consider for the sake of example the following binary systems: NS-WD with masses $m_p=1.3 M_\odot$ and $m_c=0.2 M_\odot$, NS-NS with masses $m_p = m_c=1.3 M_\odot$ and WD-WD with masses $m_p=0.5 M_\odot$ and $m_c=0.25 M_\odot$. According to our above discussion of the DM wind, and considering the observed orbital period range and binary positions, we perform our analysis varying the parameters in the following ranges: $10$~km~s$^{-1}$ $\lesssim v_{w}\lesssim 1000$~km~s$^{-1}$, $0.1$~days $\lesssim P_{b} \lesssim 100$~days and a scale radius $0.1$~kpc $\lesssim r \lesssim 10$~kpc. It is important to note that we will also consider binary systems near the Galactic center (at parsec scales) since it is of interest to check the DMDF in regions along the Galaxy where DM is supposed to dominate.

\subsection{DMDF in observed binaries}\label{sec:5a}

We first apply the approach to the Galactic binaries with measured intrinsic orbital periods and which are remarkably well explained by GW emission. In the last three columns of Table \ref{tab:pulsars} we compare $\dot{P}^{\rm GW}_b$ with $\dot{P}^{\rm DF}_b$. In this calculation we use the NFW profile and the RAR model for illustrative purposes and the following free parameters: $\beta=0$ and $v_w=100$~km~s$^{-1}$. For other values of $\beta$, $\dot{P}_{b}^{\rm DF}$ does not change significantly, however a change of $v_w$ by one order of magnitude may be more important in the computation of $\dot{P}_{b}^{\rm DF}$ as we shall see below. At this point we should discuss whether the binaries of Table \ref{tab:pulsars} are bound or unbound to the Galactic gravitational potential to determine a more precise value for the DM velocity wind\footnote{It is important to clarify that the pulsar B2127+11C is located within the Galactic globular cluster M15 whereby it is subjected dominantly to the gravitational potential of its host globular cluster. As in the case of bulge globular clusters accelerating (possibly) pulsars through their stellar components \cite{2016arXiv161204395P}, DM can also contribute to the total acceleration by the studied effect in this paper. This latter claim is motivated by recent observational analysis that point out favorably the importance of the DM component in the dynamical of globular clusters \cite{2013MNRAS.428.3648I,2016MNRAS.462.1937S}.\label{footnote 1}}. However, for the binaries of Table \ref{tab:pulsars} which are characterized by short orbital periods, we checked that this is not relevant since for any DM wind in the range $10$~km~s$^{-1}$ $\lesssim v_w\lesssim 1000$~km~s$^{-1}$ the value of $\dot{P}_{b}^{\rm DF}$ is still very small compared with the $\dot{P}_{b}^{\rm GW}$ and with the measured intrinsic orbital period decay.

As we can see the DMDF effect is very small for all the above binaries because of the short orbital periods (compact orbits) that lead them to experience a small drag force. 

We can thus first conclude that, for the binary systems listed in Table \ref{tab:pulsars}, the DMDF effect is indeed negligible and their secular evolution is fully dominated by GW emission.

\subsection{DMDF as a function of the orbital period}\label{sec:5b}

\begin{figure}
\centering
\includegraphics[width=1.0\hsize,clip]{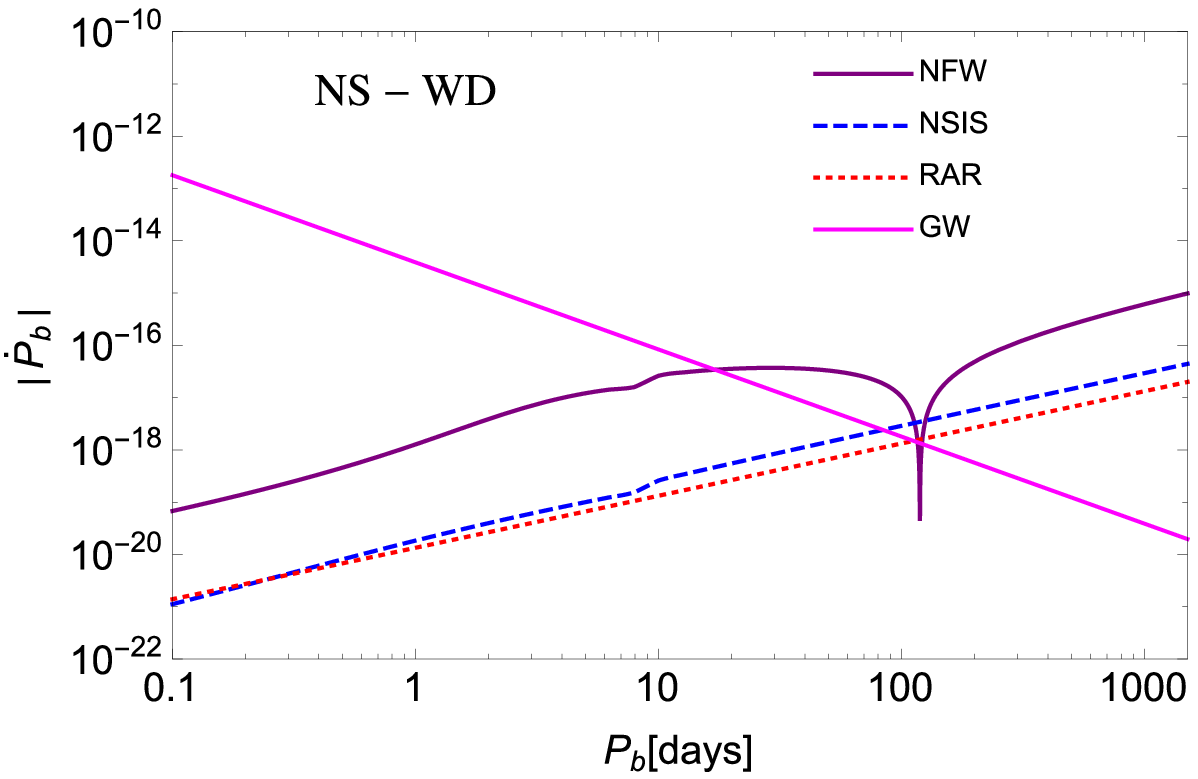}
\includegraphics[width=1.0\hsize,clip]{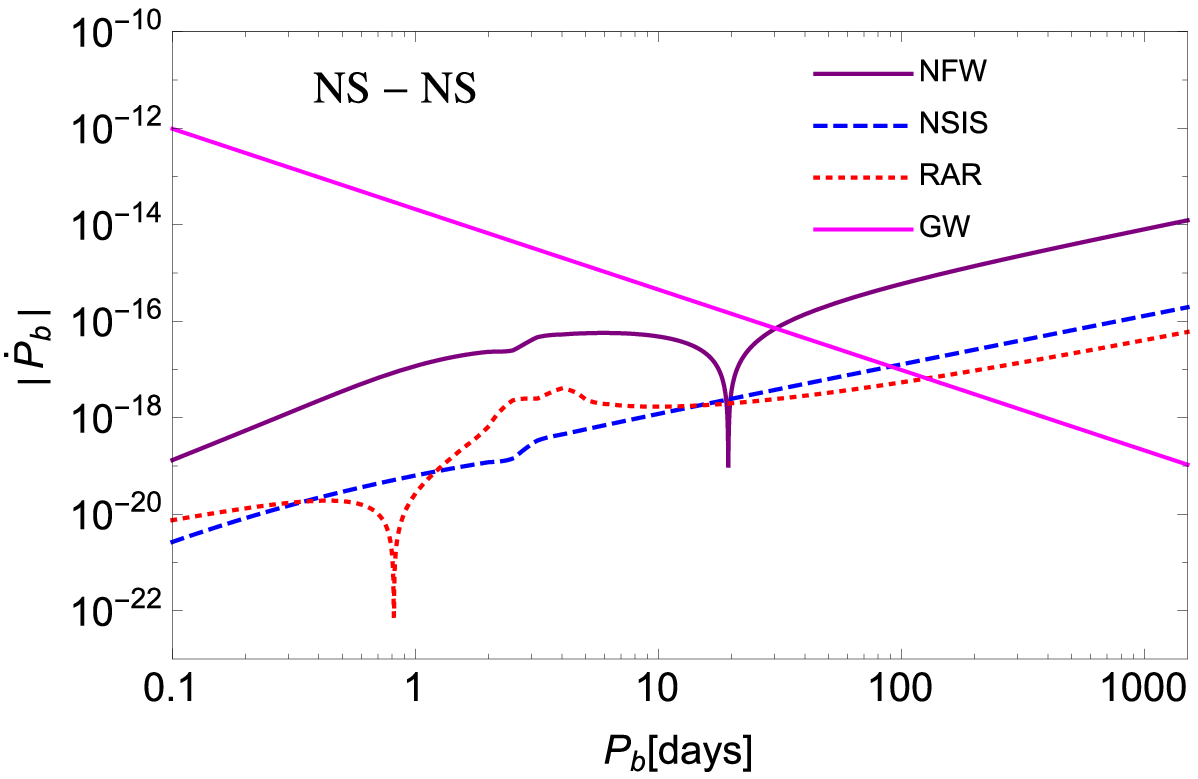}
\includegraphics[width=1.0\hsize,clip]{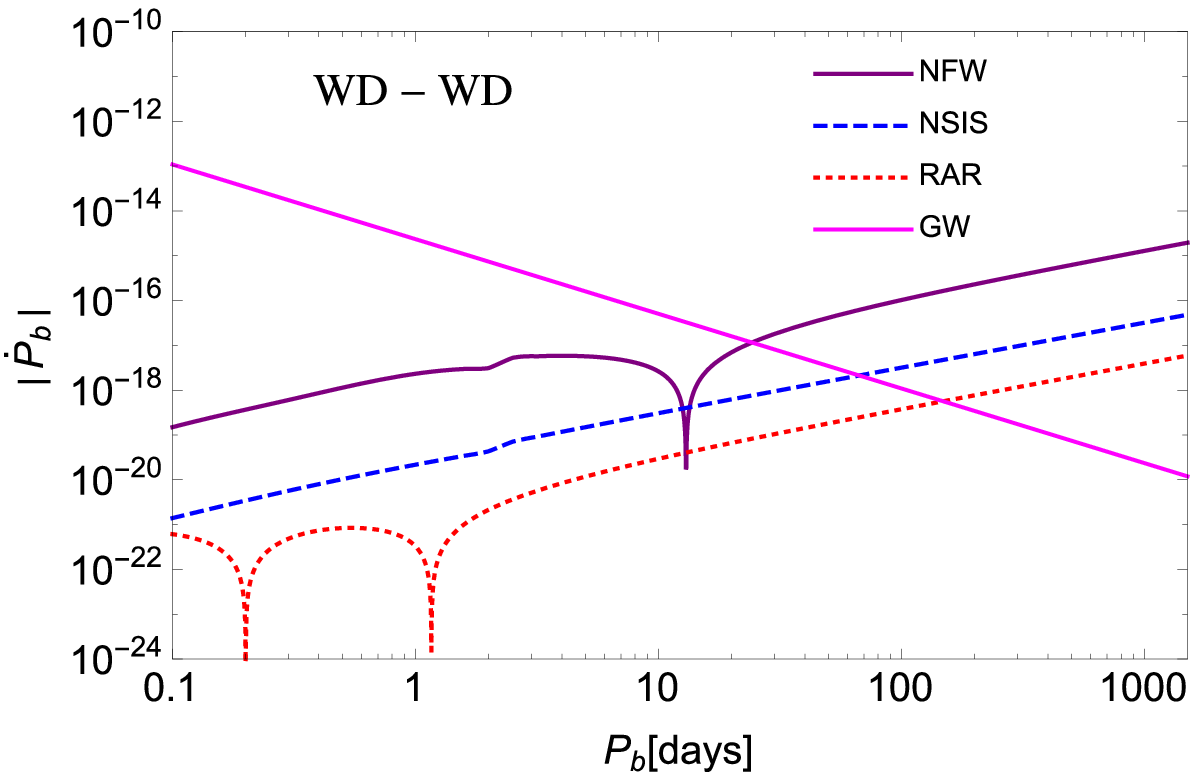}
\caption{Secular change of the orbital period as a function of the orbital period. The red dotted curve refers to the most compact solution of the RAR profile for the Milky Way, namely for a DM particle mass $m=345$~keV. The blue dashed curve shows the results for the NSIS profile and the purple solid curve the ones for the NFW profile. The pink solid line shows the prediction of the orbital decay due to GW emission. We have here adopted the values $r=0.1$~kpc, $v_w=100$~km~s$^{-1}$ and $\beta=\pi/2$. Top panel: NS-WD with $m_p=1.3~M_\odot$ and $m_c=0.2~M_\odot$. Middle panel: NS-NS with $m_p=m_c=1.3~M_\odot$. Bottom panel: WD-WD with $m_p=0.5~M_\odot$ and $m_c=0.25~M_\odot$.}\label{fig:pbd-pb}
\end{figure}

A natural question that arises is whether DMDF effects can be comparable with the orbital period decay predicted by GW emission. To answer this question we explore the physical conditions (and hence the values for the model parameters) under which such equality may be attained. We thus consider the possibility to have binary systems with large periods, e.g. $P_b=100$~days, since DMDF is enhanced in systems with small binary compactness. We also consider regions along the Galaxy where the DM is supposed to dominate as those near the Galactic center.

We start our analysis by plotting the secular change of $P_b$ as a function of the orbital period for different density profiles in Fig.~\ref{fig:pbd-pb} with values for the free parameters $v_w=100$~km~s$^{-1}$ and $r=0.1$~kpc in this analysis. We also show in the same plot the orbital decay due to GW emission $\dot{P}_{b}^{\rm GW}$, according to Eq.~(\ref{c9}).

\begin{table}
\begin{tabular}{lccc}
\hline
DM Profile & NS-WD & NS-NS & WD-WD \\ \hline
NFW & 18 & 30 & 25 \\ 
RAR & 120 & 130 & 150 \\ 
NSIS & 80 & 90 & 70 \\
\hline
\end{tabular}
\caption{This table displays theoretical predictions of orbital periods in days at which $\dot{P}_{b}^{\rm DF}$, computed by the indicated DM density profiles, equates $\dot{P}_b^{\rm GW}$ predicted by general relativity for different binary systems.}\label{tab:periods}
\end{table}

We can see that for a NS-WD system (top panel in Fig.~\ref{fig:pbd-pb}), the orbital period decay starts to be dominated by the DMDF effect shortly after than $P_b=18$~days, for the NFW profile. i.e  it is now larger that the one predicted by the GW emission. For the same system, The NSIS predicts a $\dot{P}_{b}^{DF}$ that matches $\dot{P}_{b}^{\rm GW}$ around $P_b=80$~days while for the RAR model, it occurs around $120$~days. For a NS-NS system (middle panel in Fig.~\ref{fig:pbd-pb}), the NFW provides the match around $30$~days and around $90$~days and $130$~days for the NSIS and RAR model respectively. For a WD-WD system (bottom panel in Fig.~\ref{fig:pbd-pb}), the NFW provides the match around $25$~days and around $70$~days and $150$~days for the NSIS and RAR model respectively. These results are also summarized in Table \ref{tab:periods} for clarity. However, for such large periods, DMDF provides small orbital decays between $10^{-16}$ (for the NFW profile) and around $10^{-18}$ (for the other profiles) as can be seen in Fig.~\ref{fig:pbd-pb}. It is evident from here that, the larger the orbital period, the larger the $\dot{P}_{b}$ reached. For instance for $P_{b}=1000$~days, $\dot{P}_{b}\sim 10^{-14}$ for the NFW profile and NS-NS binaries (middle panel in Fig.~\ref{fig:pbd-pb}). These values are however very small, with respect, for instance, to the measured intrinsic orbital decays shown in Table~\ref{tab:pulsars} for some binary systems. However, possible measurements of the intrinsic period decays for binary systems with characteristic large periods is a challenge of unprecedented precision for astronomical observations. If such measurements might be successfully attained, it could also lead to discriminate between different DM density profiles due to the outstanding precision which is a characteristic property in such systems.

\subsection{DMDF as a function of the DM wind}\label{sec:5c}

\begin{figure}
\centering
\includegraphics[width=1.0\hsize,clip]{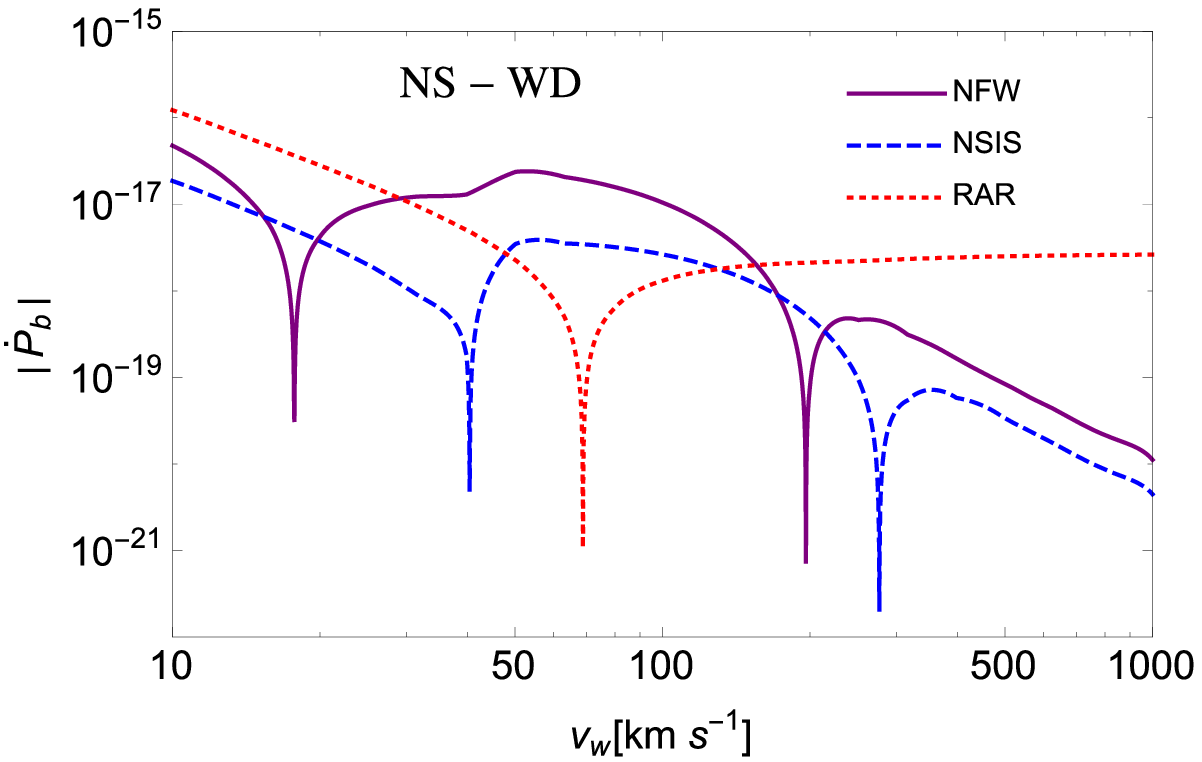}
\includegraphics[width=1.0\hsize,clip]{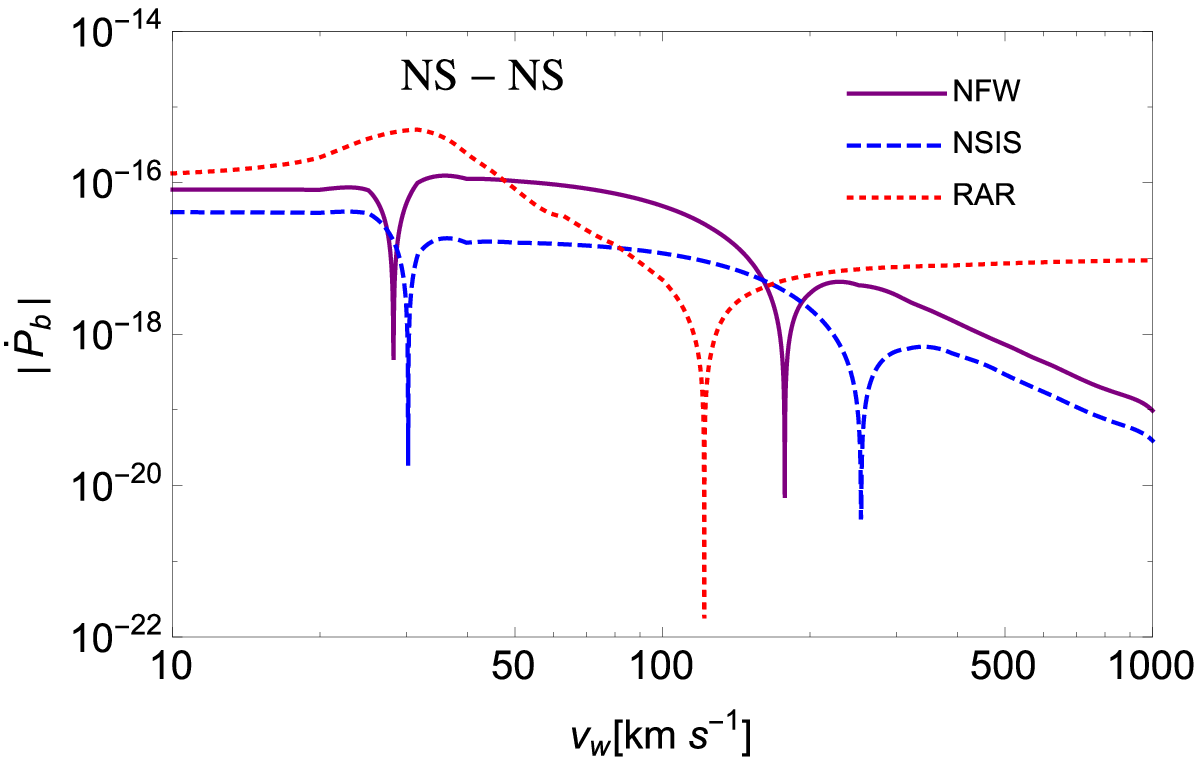}
\includegraphics[width=1.0\hsize,clip]{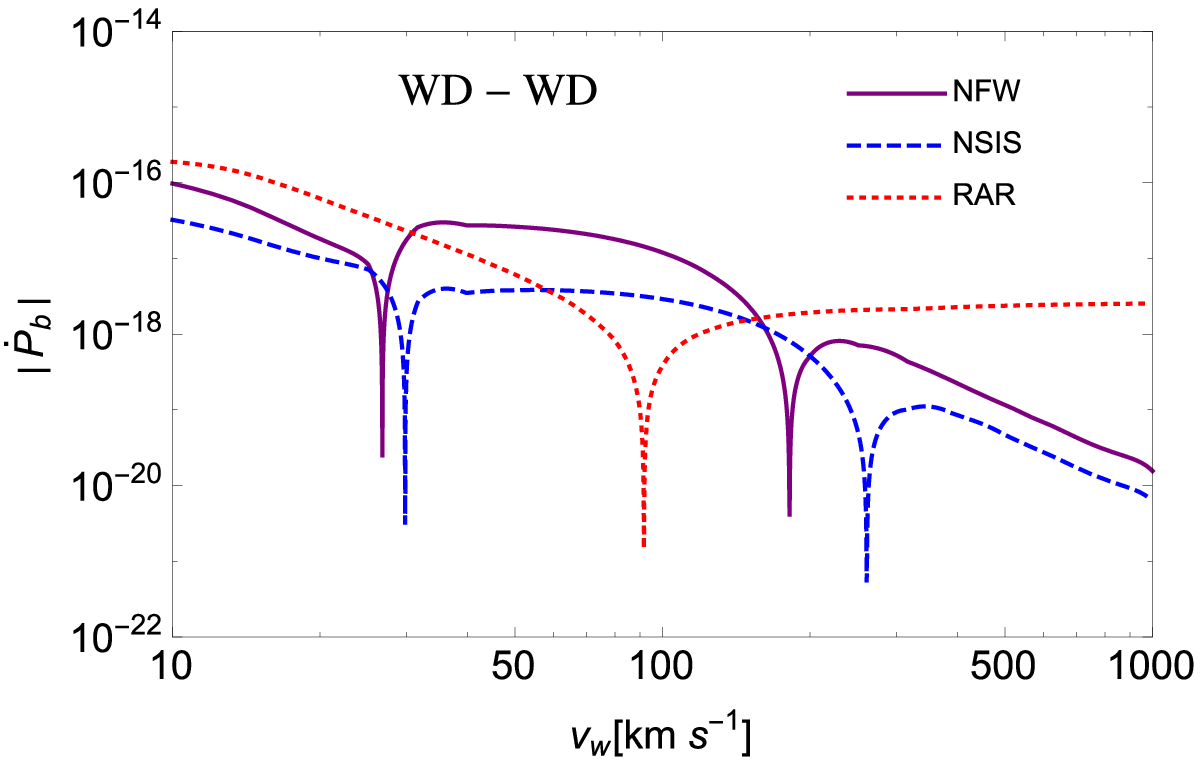}
\caption{Secular change of the orbital period as a function of the DM velocity wind. The red dotted curve refers to the most compact solution of the RAR profile for the Milky Way, namely for a DM particle mass $m=345$~keV. The blue dashed curve shows the results for the NSIS profile and the purple solid curve the ones for the NFW profile. The pink solid line shows the prediction of the orbital decay due to GW emission. We have here adopted the values $r=1.5$~kpc, $P_b=100$~days and $\beta=\pi/2$. Top panel: NS-WD with $m_p=1.3~M_\odot$ and $m_c=0.2~M_\odot$. Middle panel: NS-NS with $m_p=m_c=1.3~M_\odot$. Bottom panel: WD-WD with $m_p=0.5~M_\odot$ and $m_c=0.25~M_\odot$.}\label{fig:pbd-vw}
\end{figure}
In order to analyze the effect of the wind velocity, we choose the radial position of the binary system fixed (measured from the Galactic center) at $r=1.5$~kpc and the orbital period $P_b=100$~days. The Fig.~\ref{fig:pbd-vw} shows that, for the aforementioned parameters and for the NFW and the NSIS profile, $\dot{P}^{\rm DF}_{b}$ lies in the range $10^{-20}$--$10^{-16}$. We can see from here that the smaller the DM wind velocity the larger the $\dot{P}^{\rm DF}_{b}$. However the latter statement does not apply for the RAR model which exhibits a constant value of $\dot{P}_{b}^{DF}\sim 5\times 10^{-18}$ for NS-WD and WD-WD and around $10^{-17}$ for NS-NS, for $v_w \gtrsim 200$~km s$^{-1}$. this analysis leads to conclude that binaries into a DM background with small DM wind velocities (than the orbital velocity), experience a more effective drag force and hence a larger $\dot{P}_{b}$. We shall be then more interested in binary systems with small wind velocities, however we do not exclude at all binaries with (at least) one NS companion which may posse high kick velocities and then large wind velocities.

\subsection{DMDF as a function of the binary position}\label{sec:5d}

\begin{figure*}
\includegraphics[width=0.49\hsize,clip]{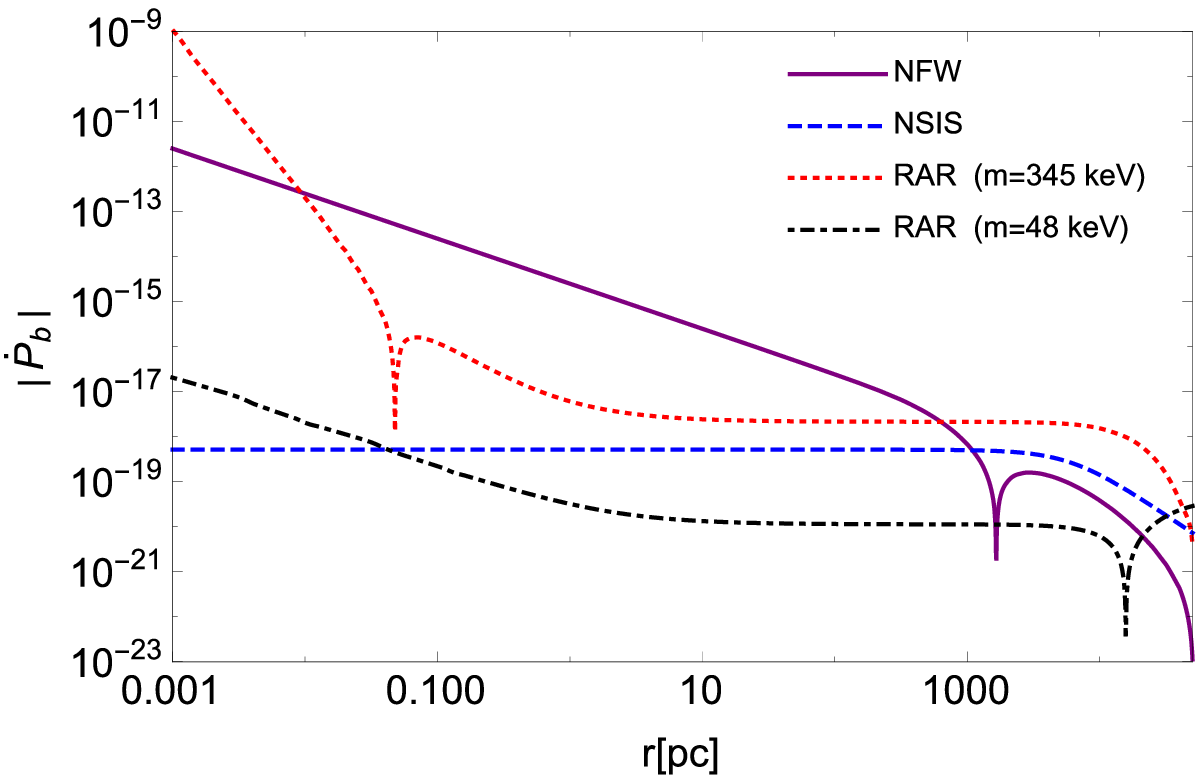}
\includegraphics[width=0.49\hsize,clip]{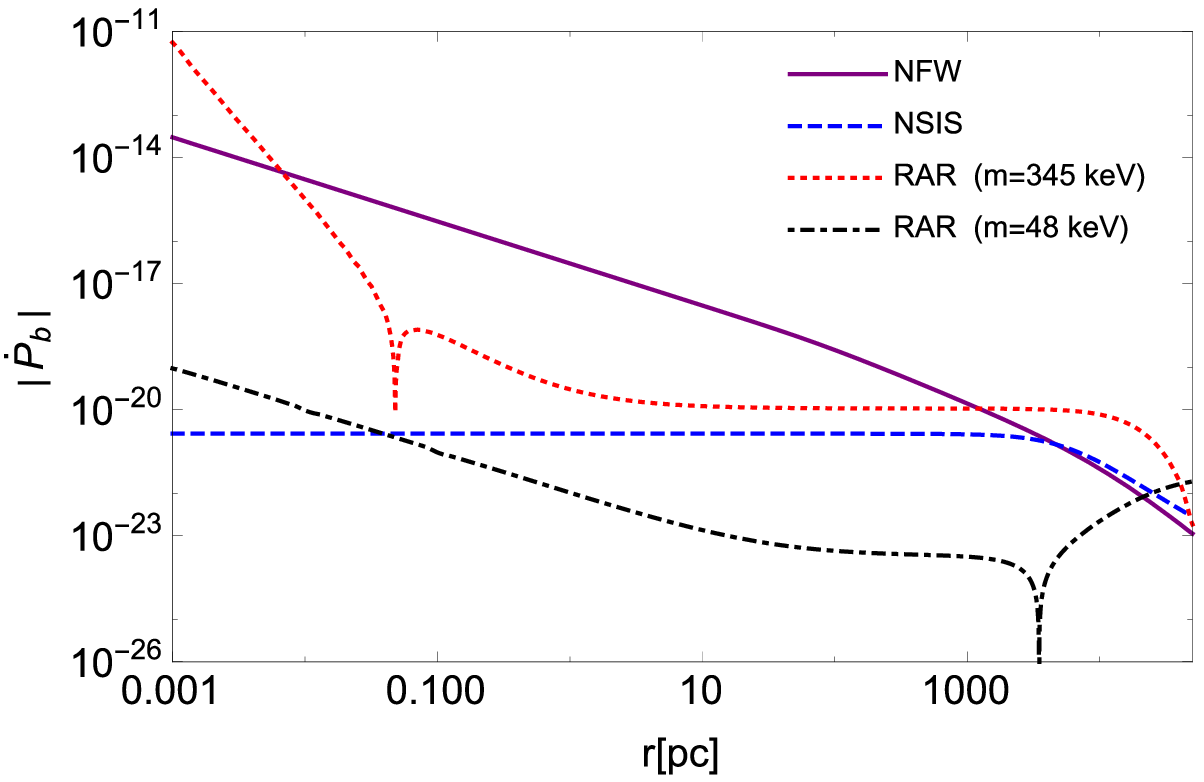}
\caption{Secular change of the orbital period of a NS-WD as a function of the radial position, for all the density profiles analyzed in this work. The red dotted curve refers to the most compact solution of the RAR profile for the Milky Way, namely for a DM particle mass $m=345$~keV. The black dot-dashed curve shows the RAR profile for $m=48$~keV. The blue dashed curve shows the results for the NSIS profile and the purple solid curve the ones for the NFW profile. We have here adopted the values $v_{w}=200$~km~s$^{-1}$ and $\beta=\pi/2$. Left panel: numerical results for the case $P_b=100$~days. Right panel: numerical results for the case $P_{b}=0.5$~days.}\label{fig:pbd-r}
\end{figure*}

We turn now to plot in Fig.~\ref{fig:pbd-r} the value of $\dot{P}_b$ as a function of the radial position. We here adopt for the DM wind $v_w=200$~km~s$^{-1}$ and for the binary period $P_b=100$~days (left panel) and $P_{b}=0.5$~days (right panel).  We can see that, differences between the solution provided by $m=48$~keV and the one provided by $m=345$~keV for the RAR model is $\approx 3\times 10^{3}$. Interestingly, towards the Galactic center, for the two chosen cases of orbital periods ($P_{b}=0.5$, 100~days), the NFW profile and the RAR model (for the most compact solution $m=345$~keV) can reach a value of $\dot{P}_{b}$ that may be comparable with the one provided by Eq.~(\ref{c9}) due to GW emissions. The prediction of $\dot{P}_{b}$ due to DMDF, for binaries with short orbital periods, can also be seen in Table.~\ref{tab:pulsars} for the NFW profile and the RAR model, respectively. For large periods however (left panel in Fig.~\ref{fig:pbd-r}), DMDF effect is highly enhanced for all the binary systems as can be seen in Fig.~\ref{fig:pbd-pb}. In particular, the RAR model predicts large orbital period decay very near the Galactic center (around $10^{-3}$~pc) due to the high DM density at such distances (see also Fig.~\ref{density}). 
The most promising situation arises then for binary positions near the Galactic center either for long or short orbital periods. We expect hence that observational measurements reach a technological improvement that permit us to measure such short orbital periods decays with outstanding precision in the future. In addition, it would be interesting to observe binary systems near the Galactic center to put constraints on the Galactic center environment, particularly on the DM density profile and importantly, to check the GR predictions in the strong field regime. 

\subsection{From orbital shrinking to widening}\label{sec:5e}

%
\begin{figure}
\centering
\includegraphics[width=\hsize,clip]{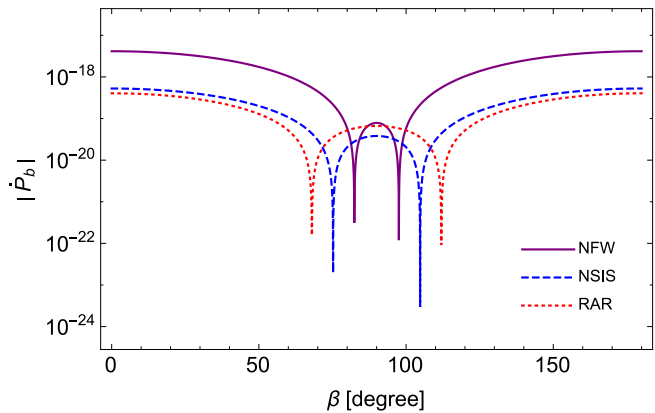}
\caption{Secular change of the orbital period for a NS-WD as a function of the angle $\beta$. The red-dotted curve refers to the most compact solution of the RAR profile for the Milky Way, namely for a DM particle mass $m=345$~keV. The blue-dashed curve shows the results for the NSIS profile and the purple-solid curve the one for the NFW profile. We have here adopted the values $P_b=100$~days and $r=1.5$~kpc for all the profiles. Here we adopt values of the wind velocities that can lead to change of sing in the orbital period time-derivative. For the RAR model $v_w=70$~km~s$^{-1}$, for NFW profile $v_w=200$~km~s$^{-1}$ and for the NSIS profile $v_w=300$~km~s$^{-1}$.} \label{pbd-beta}
\end{figure}

We turn now to analyze the model parameters under which a change of sign in the orbital period first time-derivative occurs. Namely, the conditions under which DMDF produces an orbital widening instead of an orbital shrinking or vice-versa. For given binary parameters and $\beta$, there are values of the wind velocity for which occurs a change of sign of $\dot{P}_b$. This is clearly seen in Fig.~\ref{fig:pbd-vw} for each density profile, for binaries with known values of the orbital period and distance and setting $\beta = \pi/2$. Fig.~\ref{pbd-beta} shows, instead, how sensitive is this feature to the value of the $\beta$ parameter and to the Galactic DM distribution, i.e on the DM density profile. 
We describe now, for the sake of example and without loss of generality, the case of the NS-WD binary of Fig.~\ref{pbd-beta}. In this analysis we have adopted $P_b=100$~days and $r=1.5$~kpc as known quantities. The two changes of signs occur at: $\beta = 68.75\degree$ and $114.6\degree$ for the RAR model with $v_w=70$~km~s$^{-1}$; $\beta = 80.21\degree$ and $97.40\degree$ for the NFW profile with $v_w=200$~km~s$^{-1}$; and $\beta=74.49\degree$ and $103.13\degree$ for the NSIS profile  with $v_w=300$~km~s$^{-1}$ (see Fig.~\ref{pbd-beta}). These results are in general agreement with the ones found in Ref.~\cite{2015PhRvD..92l3530P} within the approximation of large $v_w$. The contribution of fast moving particles with respect to the binary-components, along with particular choices of $v_w$, $\beta$ and even $P_{b}$, might lead (although it is not a necessary condition) to multiple changes of sign of $\dot{P}_b$. This analysis supports the necessity of taking into account this contribution to check the conditions under which $\dot{P}_b$ may change sign. If one were interested in providing only negative values of $\dot{P}_b$, the particular choice $\beta=0$ (or more generally a value of it out the above ranges) would fulfill such requirement. We set then henceforth on the contrary $\beta=\pi/2$ in order to introduce a possible change of sign in $\dot{P}_b$ as a general case.

Let us turn back to Fig.~\ref{fig:pbd-pb}. We note that the change of sign can occur for shorter or longer periods depending on the density profile and the DM wind velocity. For instance, for the NFW profile and $v_w=200$~km~s$^{-1}$, the change of sing occurs at $P_b\approx 2$~days contrary to the case $v_w=100$~km~s$^{-1}$ where the change is around $100$~days for NS-WD as can be seen in the top panel of Fig.~\ref{fig:pbd-pb}. The change of sign of $\dot{P}_b$ may occur at short period values for the RAR model, around 1~day for NS-NS and WD-WD (see middle and bottom panel in Fig.~\ref{fig:pbd-pb}), while for the NFW profile, it may occurs around 20~days for NS-NS, around 12~days for WD-WD and around 100~days for NS-WD. The NSIS profile always provide negatives values in all the binary systems shown in Fig.~\ref{fig:pbd-pb}. Before the first peak and after the second one, $\dot{P}_b$ is always negative while between the two peaks $\dot{P}_b$ is positive. We recall that, however, negatives values can be obtained for all the orbital period range in the case in which $\beta$ takes a value different from the aforementioned ranges, independently of the binary system and the other parameters as was inferred from Fig.~\ref{pbd-beta}.

It can be seen that negative values of $\dot{P}_b$ correspond to those binaries between the two peaks contrary to the curve given by the RAR model in the bottom panel in Fig~\ref{fig:pbd-pb}. We also note that the position of those peaks does not change significantly when the orbital period varies, but rather the order of magnitude of $\dot{P}_b$. In some cases it can vary up to one order of magnitude. It is also important to note that this feature may change depending on the radial position and the density profile. As we can see from the same plot, the RAR model shows negative values of $\dot{P}_b$ below the first peak.

We can analyze the behavior as a function of the binary position in Fig.~\ref{fig:pbd-r}. For the case of NFW profile and left panel ($P_b=100$~days), negatives values of $\dot{P}_{b}$ can still be found after the peak as pointed out previously; therefore, positives values are located below $1.5$~kpc for the NFW profile, while for both the NSIS profile and the RAR model, $\dot{P}_{b}$ is always negative. In the right panel of the same figure (for $P_b=0.5$~days), all the DM density profiles provide negatives values of $\dot{P}_b$ except the RAR model, before the peak, for $m=48$~keV (also for $P_b=100$~days). It is important to stress that this analysis is valid for $\beta=\pi/2$ since, for other values of it, as $\beta=0$, $\dot{P}_b$ is always negative being independent of the DM density profile as can be inferred from Fig.~\ref{pbd-beta}.

\section{Discussion and conclusions}\label{sec:6}

It is by now well-known that the high-precision measurements of the orbital parameters of compact-star binaries (e.g. NS-NS, NS-WD and WD-WD) with short orbital periods ($P_b\lesssim 0.1$~days) have allowed a remarkable verification of the of the orbital decay predicted by general relativity due to GW emission (see Table \ref{tab:periods} and references therein). However, the binary gravitational binding energy can be also affected by an usually neglected phenomenon, namely the DMDF (i.e. DM gravitational drag) induced by the DM on the binary owing to the interaction of the binary components with their DM gravitational wakes. We have qualified and quantified in this work this effect in the evolution of compact-star binaries and assessed the conditions under which it can become comparable to the one of the GW emission. We can draw the following conclusions from such an analysis:

\begin{enumerate}

\item 
A first interesting situation may occur for binaries with long orbital periods above 20~days: the orbital decay produced by DMDF becomes comparable to the one produced by the emission of GWs. Clearly, the precise orbital period at which the two effects are quantitatively equal depends on the DM density profile and on the binary parameters (see Fig.~\ref{fig:pbd-pb}). 

\item 
We have presented here, for the NFW, the NSIS and the RAR DM profiles, the orbital period for NS-NS, NS-WD and WD-WD binaries at which the DMDF effects, start to dominate over the produced by GW emission. These results are summarized in Table \ref{tab:periods} (see also Fig.~\ref{fig:pbd-pb}). 

\item 
The NFW profile and the RAR model provide a more significant effect in the drag force than the one given by the NSIS profile, as can be seen in Figs.~\ref{fig:pbd-pb}--\ref{fig:pbd-r}. It is important to note that the RAR and the NSIS profile predictions are similar above $P_b = 100$~days, for all the binary systems analyzed in this work (with $v_w=100$~km~s$^{-1}$ and located at $0.1$~kpc), while also for those values of $P_b$ the NFW profile predicts a much larger DMDF effect.

\item 
Another promising situation arises for binary systems located very near the Galactic center. In this case, the $\dot{P}_{b}$ due to DMDF is increased even for short orbital periods ($P_{b}=0.5$~days) as is shown in the  right panel of Fig.~\ref{fig:pbd-r}. For long orbital periods the DMDF is notoriously strengthened, particularly for the NFW profile and the most compact solution for the RAR model ($m=345$~keV). This latter situation corresponds to the most ideal case for testing the DMDF (left panel of Fig.~\ref{fig:pbd-r}). 

\item For the most ideal scenario of the DMDF effects in binary systems, kinematic effects, which are proportional to the orbital period, must be considered and respectively compared to the one studied in this work. 

\item It is known that positive values of $\dot{P}_{b}$ can be caused for example by binary mass-loss or mass-exchange. However, we have seen that $\dot{P}_{b}$ might change sign from negative to positive due to DMDF. This is shown in Fig.~\ref{pbd-beta} for different  DM density profiles. Thus, this effect could be study in binary systems dominated by kinematic effects.

\end{enumerate}

To summarize, The DMDF is very sensitive to the DM properties: density profile, velocity distribution function and velocity dispersion profile; whereby it would permit to put stringent constraints on the DM properties (and presumably on the nature) at the binary position and thus to discriminate between different DM models. Following this idea, the determination of the orbital secular changes of compact-star binaries with long/short orbital periods located in the outer halo/center of the Galaxy, might constrain the DM density distribution in these locations. It would be also interesting to study such an effect in binaries with measured orbital decays within globular clusters (as in the case of B2127+11C) in order to put constraints on the DM distribution in these systems. Therefore, the possible identification of this effect establishes a topic for future high-precision astrophysical data for the analysis of the secular evolution of compact-star binaries.


\section{Acknowledgments}
L.G.G. is supported by the Erasmus Mundus Joint Doctorate Program by Grant Number 2013--1471 from  the EACEA  of the European Commission. J.A.R. acknowledges the partial support of the project N 3101/GF4 IPC-11, and the target program F.0679 of the Ministry of Education and Science of the Republic of Kazakhstan. It is a pleasure to thank C.~R.~Argu\"elles for interesting comments and discussions on the subject of this work.


\appendix*

\section{Distribution functions from the Eddington's formula}

For a given density profile, the gravitational potential can be obtained by solving the Poisson's equation $\nabla \Psi=-4\pi\rho(r)$. Now, in order to solve the Eddington's formula Eq.~(\ref{a4}), we express the integral there in terms of $r$ instead of $\Psi$, and choose the appropriate limits of integrations by inverting numerically the equation $\Psi(r)=g(r)$, with $g(r)$ being a defined function of the radial position for a given density profile. In addition, the condition that the distribution function be positive for any positive energy, i.e, $f(\mathcal{E})\ge 0$ for $\mathcal{E}\ge 0$, should be guaranteed. This condition is fulfilled when $\Phi(r)$ goes to zero at infinity along with the appropriated value of the central potential $\Phi_{0}=\Phi(r=0)$.
\begin{figure}
\centering
\includegraphics[width=\hsize,clip]{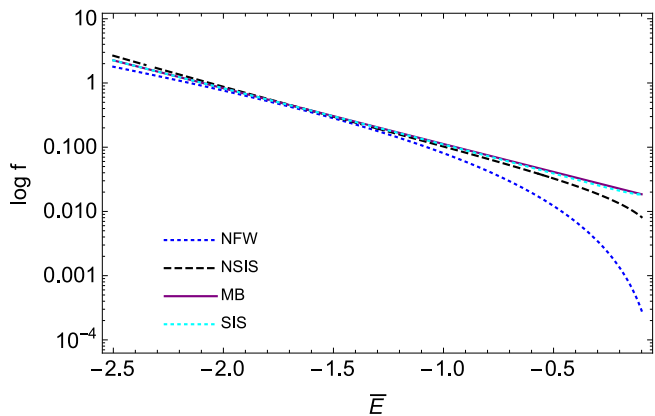}
\caption{Distribution functions for all the density profiles listed in the legend. These self-consistent distributions correspond to the solution of the Eddington's formula Eq.~(\ref{a4}).} \label{Eddign_sol}
\end{figure}
We then set $\mathcal{E}=-E$ and do for convenience the simple change
\begin{equation}
\frac{d^{2}\bar{\rho}}{d\bar{\Psi}^{2}}=\frac{d}{d\bar{r}}\left( \frac{\bar{\rho}\prime(\bar{r})}{\bar{\Psi}\prime (\bar{r})}\right)\frac{d\bar{r}}{d\bar{\Psi}}.
\end{equation}
All quantities with bar are dimensionless by making use of the model parameters of the respective density profile. With all of this, we can perform numerically the integral in Eq.~(\ref{a4}). In order to compare the distribution functions associated with the NFW and the NSIS profiles, with the Maxwell-Boltzmann one, we have to normalize $f$ to common units. For this we follow the usual normalization $\sqrt{8} M/(R V)^{3}$, where $M$ is the mass enclosed at a position $R$ where the circular velocity $V$ becomes flat, and $E$ is given in units of square velocity $V^{2}$; hence we introduce the dimensionless quantity $\bar{E}=E/V^{2}$. For the NFW profile (Eq.~\ref{a1}) such a radius is given by the virial radius, $r_{\rm v}=c r_{\rm s}$, where $c$ is the so-called concentration parameter $c$ and $r_{\rm s}$ is the scale radius. For this profile, we measure then $f$ in units of $\sqrt{8} M_{\rm v}/(R_{\rm v} V_{\rm v})^{3}$ and $\bar{E}=E/V_{\rm v}^{2}$. For the NSIS profile (Eq.~\ref{a2}), we adopt the core radius $r_0$ and thus all the quantities derived from it such that $f$ is given in units of $\sqrt{8} M_{0}/(r_{0}V_{0})^{3}$ and $\bar{E}=E/V_{0}^{2}$. Furthermore, we express the Maxwell-Boltzmann distribution as follows \cite{2008gady.book.....B} 
\begin{equation}
f(\mathcal{E})=\frac{\bar{\rho}_{0}}{(2\pi\sigma^{2})^{3/2}} \exp{[\mathcal{E}/\sigma^{2}]},
\end{equation}
with $\bar{E}=-\mathcal{E}/2\sigma^{2}$. 

To check the consistency of our calculation we also apply the above method to the singular isothermal sphere (SIS)
\begin{equation}
\rho(r)=\frac{\sigma^{2}}{2\pi G \bar{r}_{0}^2}\left(\frac{\bar{r}_{0}}{r}\right)^{2},
\end{equation}
which must follow the the Maxwell-Boltzmann distribution function (see, e.g., \cite{2008gady.book.....B}). We define the central density $\bar{\rho}_{0}=\sigma^{2}/2\pi G \bar{r}_{0}^{2}$ and compute its associated distribution function also from the Eddington's formula. This solution is represented by the cyan-dotted line in Fig.~\ref{Eddign_sol} which can be seen overlaps with the Maxwell-Boltzmann distribution. For this profile, we measure $f$ in units of $\sqrt{8} \bar{M}_{0}/(\bar{r}_{0} \sqrt{2}\sigma)^{3}$. Thus, all the distribution functions and the dimensionless energy $\bar{E}$ are given in terms of theirs model parameters. Therefore, once we set the units of $f$ and $\bar{E}$, we can infer quantitatively the scale factor that lead to compare our results. However, it is important to mention that such a scale factor may be somewhat arbitrary when one does not consider a finite size for the halo which forces to introduce a cutoff at some radius scale. The relation between the relative energy $\mathcal{E}=-E$ and the particle velocity $v$ is determined by $E=\frac{1}{2}(v^{2}-v_{esc}^2)$. For $v<v_{esc}$ particles are of course bounded. Finally, we present numerical results of the distribution functions associated to the NFW and the NSIS profiles and the comparison with the Maxwell-Boltzmann distribution function in Fig.~\ref{Eddign_sol}. Our goal in this computation is to validate the approximation of taking the Maxwell-Boltzmann distribution to describe the velocity distribution for the aforementioned profiles. We can see that, the largest differences occur close to unbound energies, where precisely the contribution of particle velocities near the escape velocity do not contribute significantly to the dynamical friction force. These results then lead us to approximate, within our estimations, the velocity distributions function for the aforementioned profiles to follow the Maxwell-Boltzmann distribution. Such approximation permits to facilitate notoriously all the numerical computations regarding the orbital period decay. However, if we had at disposition observational timing pulsar data to test robustly our predictions, we would have to use the exact velocity distribution function for every density profile according to the Eddington's formula. 



%

\end{document}